\documentstyle [prl,aps,preprint]{revtex}

\def\pf6{(TMTSF)$_2$PF$_6$}
\def\clo4{(TMTSF)$_2$ClO$_4$}

\begin{document}
\draft


\title{
Quantum Hall Transitions in (TMTSF)$_2$PF$_6$}

\author{S. Valfells,$^a$ J.S. Brooks,$^{a,b}$ Z. Wang$^a$}
\address{$^a$ Department of Physics, Boston University, Boston, MA 
02215}
\address{$^b$ Department of Physics, Florida State University, 
Tallahassee, FL 32306-4005}
\author{S. Takasaki,$^c$
J. Yamada,$^c$ H. Anzai,$^c$ and M. Tokumoto$^d$}
\address{$^c$ Himeji Institute of Technology, 2167 Shosya, Himeji, 
Hyogo 671-22, Japan}
\address{$^d$ Electrotechnical Laboratory, Tsukuba, Ibaraki 305 , 
Japan}

\date{\today}
\maketitle
\mediumtext
\begin{abstract}

We have studied the temperature dependence of the integer 
quantum Hall 
transitions in the molecular crystal \pf6. We find that the transition width
between the quantum Hall plateaus does not exhibit the universal 
power-law scaling behavior of the integer quantum Hall effect 
observed 
in semiconducting devices. Instead, the slope of the $\rho_{xy}$ 
risers, $d\rho_{xy}/dB$, and the (inverse) width of the $\rho_{xx}$
peaks, $(\Delta B)^{-1}$, show a BCS-like energy gap  temperature 
dependence.
We discuss these results in terms of the field-induced spin-density 
wave
gap and order parameter of the system.\\

PACS numbers: 72.80.L, 73.40.H, 75.30.F\\

\end{abstract}
\narrowtext
\newpage

One of the key features of the quantum Hall effect (QHE) in two 
dimensional semiconductor heterostructures is the universal 
scaling behavior
of the transitions between adjacent quantized Hall plateaus
\cite{wei1}.
For the integer QHE, the maximum slope of the Hall step 
($d\rho_{xy}/dB)$,
and the half-width of the $\rho_{xx}$ peak ($\Delta B$)
scale as $(d\rho_{xy}/dB)^{-1} \sim \Delta B\sim T^{\kappa}$
with $\kappa \approx 0.42$ \cite{wei1}.
This scaling behavior indicates that the quantum Hall transitions
in semiconductor heterostructures are continuous zero temperature
phase transitions characterized by a single relevant length scale
$\xi_{\rm loc}$, the localization length, which diverges as
$\xi_{\rm loc} \sim |\Delta B|^{-\nu}$ in the low temperature limit.
The localization of the electronic states, the resurrection of
extended states at the center of the Landau
levels, and the existence of scaling for the quantum Hall 
transitions as the Landau levels pass through the Fermi
energy are intrinsic properties of a weakly disordered
two dimensional electron gas (2DEG) in a strong magnetic field.

Under quite different settings, the quantized Hall resistance plateau
behavior has been convincingly demonstrated
in the quasi-2D molecular crystals \pf6 \cite{jcooper,sth2}
and \clo4 \cite{qhclo4}. 
In contrast to the case of 2DEG, the
quantization of the Hall resistance in the organic crystals 
is due to the instability of the metallic open Fermi surface
against spin-density-wave formation \cite{gorkov}.
As pointed out by Poilblanc {\it et al.} \cite{poilblanc} and Azbel {\it
et al.} \cite{azbel}, the quantized Hall effect occurs as a result of
the opening of a succession of gaps in the quasiparticle spectrum
associated with the quantized nesting vectors of
a series of magnetic field induced spin density wave (FISDW) states.
The free energy of the system is at a minimum when
the Fermi energy lies in one of the FISDW gaps which separate
filled and empty Landau levels. 
Since the ordering wavevector changes discontinuously, the 
transitions
between the FISDW subphases are first order.
Consequently, the transitions between the quantized Hall plateaus
are expected to be discontinuous {\em in the absence of impurities}.
However, the precise nature and the properties of the quantum Hall
transitions in realistic materials have not been investigated.
In this Letter we address this problem through 
the measured temperature dependence of the
magnetoresistances in \pf6. We find
that $(\Delta B)^{-1}$ and
$d\rho_{xy}/dB$, defined in a similar way as in the case of 2DEG,
do not exhibit power-law scaling. Instead, they
scale qualitatively as the square of the BCS-like, FISDW energy gaps.

\pf6\cite{ishiguro,kang1} is a highly anisotropic 
conductor: parallel chains of the
TMTSF molecules form planes separated by
the PF$_{6}$ complex, with transfer integral 
values of $t_{a}$ (along the chains), $t_{b}$ 
(between adjacent, co-planar chains), and $t_{c}$ 
(between planes) approximately equal to 
$300, 30$ and $1~meV$, respectively.  
At ambient pressure, \pf6 undergoes a
spin density wave (SDW) transition at 
$T_{SDW} \approx 12~K$, but
when pressurized, it becomes 
superconducting at pressures in excess of 
$P_{c}~\approx~8~kbar$ with 
$T_{c}~\approx~1~K$\cite{pressure}.
In a magnetic field ($B \parallel$ to the $c$-axis) 
and at $P > P_{c}$, \pf6 exhibits a succession of FISDWs. 
These are explained \cite{gorkov,gilles1}
by the nesting of the slightly warped, one dimensional Fermi
surface (Fig.\ref{p_d}) through a 
succession of field dependent nesting vectors
with a quantized component along the
stacking direction: $Q_{a} = 2k_{F} + NeBb/\hbar$. 
The phase diagram of our \pf6 sample at 11 $kbar$ is shown in
Fig.\ref{p_d}. The second order transition line separates the metallic 
and the FISDW 
phases,
whereas different FISDW phases are separated by the nearly vertical
phase boundaries where the quantized Hall
plateau transitions take place. In an ideal sample without impurities,
the latter transitions are sharp and first order in nature, and the
corresponding Hall plateau transitions are discontinuous. However, in 
realistic samples where impurities are 
always
present, the transitions are broadened and their widths 
(delimited by the broken lines in Fig.\ref{p_d}) 
saturate at low temperatures.

Six gold leads were attached to the sample on the
edge perpendicular to the $b$-axis, three on each side
as shown in Fig.\ref{p_d}.
The sample was then pressurized in a BeCu cell with a fluorocarbon
fluid.  A standard AC lock-in techniques, with a current of 10$\mu 
A$,  was used to acquire the data. The Hall signal $\rho_{xy}$ 
($\rho_{xx}$ signal)
was obtained by subtracting (adding) field sweeps (at constant 
temperature)
of opposite polarity.
Results at low temperatures (50--640~$mK$) were obtained
in a dilution refrigerator and a $20~T$ superconducting magnet.  
Higher temperature measurements (0.5--4.2~$K$) were done 
in a $^{3}$He cryostat and a 30~$T$ resistive magnet.

The $\rho_{xy}$ and $\rho_{xx}$ data are shown in 
Fig.\ref{res}.
The gross features are consistent with the QHE phenomenology in 
2DEG:
Hall plateaus in $\rho_{xy}$ are evident, as are the pronounced
Shubnikov-de Haas oscillations in
$\rho_{xx}$ with minima coinciding with the location of the 
$\rho_{xy}$
plateaus. Nevertheless, there are obvious differences
\cite{jcooper,sth2}.
Since the prerequisite for observing Hall resistance quantization in
\pf6 is the opening of quasiparticle gap in the FISDW phases,
the Hall signal is zero until the sample enters the FISDW phase.
This should be contrasted to the conventional integer QHE in 2DEG 
that
requires a ``mobility gap'' provided by disorder-induced localization 
of electronic states at the Fermi energy away from the band centers.
Treating the sample as $N$ parallel 2D layers, our measurements 
yield 
a resistance value of $13.0 \pm 1.5~k\Omega$ 
per layer at the $n_H=1$ plateau. This value is consistent with the 
expected
resistance $h/2e^2$ ($12.9~k\Omega$) per layer as a result of
the unresolved spin degeneracy factor of two in the FISDW. We note 
the absence of negative  ``Ribault''
Hall steps seen in  earlier measurements
\cite{jcooper}.  Negative Hall steps seem to occur in the 
low pressure regime\cite{balicas1} ($P \approx 8$--$9~kbar$) just 
above $P_c$. Our results for 
$P \approx 
11.3 \pm
0.5~kbar$ are well above $P_c$.

The transitions have a frequency of $B_{1} = 67 \pm 7~T$, 
which is consistent with the values previously reported 
\cite{jcooper,sth2}.
The FISDW transitions are
predicted to occur at the zeros of the Bessel function 
$J_{0}(x)$, $x=2ct_{b}'/veBb$ ( $v$ is the Fermi velocity; $t_{b}'$
is the next nearest neighbor tight binding transfer energy) 
\cite{gorkov}. 
Consequently, the data can be used to estimate $t_{b}'$. 
To first approximation, 
$v$ scales as $(b)^{-1}$, and, $vb$ stays roughly 
constant when the sample is pressurized.
Using the ambient pressure values 
$v \approx 10^{7} cm/s$ and $b \approx 7.7 \AA$,
we find that $t_{b}' \approx 5~meV$ which meets the criterion
$t_{b}' \ll t_{b}$.  
Fitting our data to the ``standard model'' expression 
$B_{n_{H}} = B_{1}/(\gamma + n_{H})$ \cite{gilles2}
for the higher transitions labeled by
$n_H$ as in Fig.\ref{res} yields $\gamma = 2.9$.
The latter value corresponds to 
$t_{b}'/t_{b} \approx 10$ in good agreement with the estimate above based 
on
the value of $B_1$, especially if one takes into account the fact
that $t_{b}$ should increase with pressure \cite{gilles2}.  

In analogy to the integer quantum Hall transitions in 2DEG, 
the $\rho_{xx}$ signal (Fig.\ref{res})
peaks at the transitions between the Hall plateaus. 
We draw attention to the large amplitude of the $\rho_{xx}$ peaks 
and the small magnetoresistance background. 
The low background magnetoresistance within the Hall plateau-
FISDW phases
is indicative of a (relatively) low sample impurity content.
Indeed, as pointed out by Azbel {\it et al.} \cite{azbel},
the effects of impurities on the QHE induced by the FISDW differ 
significantly from those in the conventional integer QHE in 2DEG. 
While in the latter disorder provides localization of the electronic 
states and,
consequently, the absence of dissipation necessary for observing the 
QHE,
in the present case, impurities introduce extended states
in the FISDW gap and give rise to finite $\rho_{xx}$ and
non-integral values of the $\rho_{xy}$ plateaus.
Over the entire range of our measurements, the sample resistivity 
was in the residual resistivity limit (temperature independent) 
as shown in Fig.\ref{sc}.

We turn now to the question of how impurities affect the
transition between the quantized Hall plateaus.
Our main results are the temperature dependence of
Hall risers ($d\rho_{xy}/dB$), the half-width of the
of the $\rho_{xx}$ peaks ($\Delta B$), and the peak value
($\rho_{xx}^{\rm max}$), which are shown in Fig.\ref{sc}.
$\Delta B$ and $(d\rho_{xy}/dB)^{-1}$ independently measure the 
width of the transition between two FISDW phases of filled Landau
levels in the same manner as in the QHE in semiconductors.
Fig.\ref{sc} (a) and (b) clearly shows that they do not exhibit
power law scaling with temperature: instead, the temperature 
dependence
of each resembles more that of the BCS-like FISDW gaps
(apart from the finite slope at $T \approx T_{c}$),
and saturates to a finite value at low temperatures.
Thus, the transition width is finite
as indicated in the phase diagram Fig.~\ref{p_d}.
In fact, the low temperature data are consistent with the transitions 
taking place through certain intervening metallic phases stabilized,
presumably, by disorder. 
To a first approximation in
the transition region, let us consider the simple Drude expressions: 
$(\rho_{xy},\rho_{xx})=(B/n_{\rm eff}ec, m/n_{\rm eff}e^2\tau)$, 
where $n_{\rm eff}(T)$ is the effective carrier
concentration and $\tau$ is a temperature-independent
impurity scattering time. $d\rho_{xy}/dB$
therefore measures the T-dependence of the inverse carrier 
concentration near the transition. The dissipative resistance 
$\rho_{xx}$ in the transition
region should then follow the same temperature dependence, with 
overall
magnitude oscillating with the magnetic field.
This is supported by Fig.\ref{sc}(c) where the normalized peak values 
$\rho_{xx}^{\rm max}/\rho_0$ indeed scale approximately with 
$n_{\rm
  eff}^{-1}$

In the original work of Gor'kov and Lebed \cite{gorkov}, 
an oscillatory phase diagram between the FISDW subphases and the 
metallic state was predicted, which would imply re-entrant metallic
behavior even in pure systems.
Since the carrier concentration in the metallic
phase would be temperature independent, it should also be so in
the transition regions in the data, and the Hall signal would drop to
some constant value in these regions. Instead, the carrier 
concentration
seems to follow the SDW gap behavior for any field above the 
threshold field for the FISDW.


Our data strongly suggest that the finite transition width (within 
which there is metallic behavior)  is a consequence of the broadening 
of 
the
first order transition in ideal systems by disorder effects.
Let $w$ be the width of the transition between two adjacent 
subphases with 
gap order parameters $\Delta_{n+1}$ and 
$\Delta_{n}$. Thus  $w\sim(d\rho_{xy}/dB)^{-1}\sim \Delta B$ in the
transition region.
In the absence of impurities, the transitions between the FISDWs are
discontinuous in the thermodynamic limit.
For a finite system of linear dimension $L$, it is well known that
the rounding of the critical singularities in a first order transition
leads to a transition width that scales with $L^{-d}$ where $d$ is 
the effective dimensionality of the system \cite{fisher}. 
Denoting the coherence length by $\xi$ which remains finite at a
first order transition, then $w\propto1/(L/\xi)^d$.
In the presence of a disorder potential that couples to the FISDW
order parameter ({\it e.g.}, through lattice distortions), 
the stability of the first order transition
has to be considered. The essential physics can be captured by the
Imry-Ma argument \cite{imry}: it is energetically favorable for
the system to form domains over which there is a single phase. 
The typical size of the domains is given by the Imry-Ma length 
scale, $L_{\rm im}$,  which is determined by the impurity strength
and the detailed competition between the interfacial domain-wall 
energy
cost and the bulk energy gain. The important point is that
the free energy of the system is the sum of the free energies of 
the domains (with corrections due to the non-extensive contributions 
from
the domain wall energies), and is identified with its disorder-average
over domains of size $L_{\rm im}$. Consequently, $L_{\rm im}$ 
plays
the role of the effective finite size of the system.
The coherence length is related to the gap parameter through
$\xi \sim \hbar \bar{v}/\bar{\Delta}$, where
$\bar{\Delta}$ is an averaged gap order parameter for the two 
adjacent 
FISDW subphases ($\Delta_n$ and $\Delta_{n+1}$)
and $\bar{v}$ is an averaged Fermi velocity. The transition width
is therefore $w \propto (\hbar v / L_{\rm im} \bar{\Delta})^{d}$. 
For our quasi-2D system,  we take $d\approx2$, and $w^{-1}$
should then scale
as the square of the BCS-like FISDW gap parameters, 
which is consistent with the observed temperature dependence of
$d\rho_{xy}/dB$, $(\Delta B)^{-1}$ and $\rho_{xx}^{\rm max}/
\rho_{0}$ in Fig.\ref{sc}.

Specifically, on approaching the second order phase transition line 
between the metal and the FISDW in Fig.~\ref{p_d}, {\it i.e.}, near
$T_c$, $\Delta \approx \Delta_c(1 - T/T_{c})^{1/2}$ and
$w^{-1}  \sim 1 - T/T_{c}$, which explains the linear behavior
around $T_{c}$ in Fig.\ref{sc}.
In the low temperature limit, on the other hand, 
$\Delta \approx \Delta_{0} - (2 \pi \Delta_{0}
T)^{1/2}e^{-\Delta_{0}/kT}$, and the BCS-like gap
saturates exponentially regardless of the dimensionality; 
consequently, $w$ is limited by $L_{\rm im}$. 
Since the residual resistivity is constant over the range of 
temperatures of interest in the 
present experiment (Fig.\ref{sc}(c)), we take $L_{\rm im}$ 
to be constant.  
Consequently, as observed in Fig.\ref{sc}, 
$w^{-1}$ is constant in the low temperature limit. 
Finally, as the field sweeps away from transition region,
bulk FISDW order will take place, which is presumably assisted 
initially
by the quantum tunneling between the domains of the prevailing 
phase.

   This research is supported by NSF DMR 95-10427. Experiments 
were carried out at the National High Magnetic Field 
Laboratory (supported by the NSF and the State of Florida). 
We acknowledge A. Lacerda,  E. Palm,
T. Murphy and M. Davidson for technical assistance, and N. Bonesteel, 
C. Campos, 
L. P. Gor'kov, V. Melik-Alaverdian, P. Sandhu, and S. Uji for useful 
discussions.

\begin{figure}
\caption{Overview of \pf6. (a) Fermi surface of \pf6 
with nesting vector $Q$. (Coordinates  refer to both 
(a) and (b)).  (b) Lead
configuration of the sample. (c) Phase diagram at $11~kbar$.
$+$ mark metallic - FISDW second order phase boundary. 
$\times$, $\bowtie$, $\Diamond$, $\Box$, $\triangle$,   
and $\circ$:   Hall transitions
$n_{H}$ = 7 $\rightarrow$ 6, 6 $\rightarrow$ 5, 5 $\rightarrow$ 4,
4 $\rightarrow$ 3, 3 $\rightarrow$ 2, 2 $\rightarrow$ 1, 
respectively. The regions delimited by the broken lines 
represent the widths of the lowest two 
transitions. Open symbols and closed symbols correspond to low and high 
temperature experimental runs, respectively.}
\label{p_d}
\end{figure}

\begin{figure}
\caption{a) Hall resistivity. Main trace: complete field sweep at 50 
$mK$.  Transition steps $n_{H}=$1, 2, and 3 are indicated. Insets are 
up-sweeps only, offset from zero for clarity. 
Upper Inset: 4.2, 3.0, 2.3, 1.7, 1.4, 1.3, 
1.2, 1.1, 1.0, 0.8, 0.5~$K$.
Lower Inset: 640, 550, 400, 
300, 200, 100~$mK$. b) Longitudinal magnetoresistivity.
Main trace: complete field sweep at 50 $mK$.
Insets are up-
sweeps only, offset from zero for clarity. Sequence of temperatures is 
the same as in a). Note significant difference in hysteretic 
behavior of $\rho_{xx}$ compared with that of $\rho_{xy}$.}
\label{res}
\end{figure}

\begin{figure}
\caption{Temperature dependence of resistivity and Hall transition 
parameters. In all cases, circles mark the $2 \rightarrow 1$
transition and triangles the $3 \rightarrow 2$ transition.
Open symbols and closed symbols  correspond to low and high 
temperature experimental runs, respectively. (a) Hall transitions. 
$d\rho_{xy}/dB$ is estimated
by the slope of the riser as obtained by linear interpolation.
The broken line represents the temperature dependence of the
square of the BCS gap, $\Delta^{2}_{BCS}$, with $T_{c} = 3.0~K$
and the zero temperature gap amplitude normalized to the 
value of $d\rho_{xy}/dB$ at $T = 0~K$\protect\cite{muhl}.
(b) Magnetoresistance peak width.  
The width of the transition, $\Delta B$, is estimated 
by the full width at half maximum of a Lorentzian fit
to the peaks.
(c) Magnetoresistance peak amplitude
(normalized to the zero field residual resistivity).  
(d)  Zero field residual resistivity 
of  $\rho_{xx}$ as a function of temperature.}
\label{sc}
\end{figure}

\end{document}